\documentclass[%
superscriptaddress,
 amsmath,amssymb,
aps,
prb,
twocolumn
]{revtex4-1}

\usepackage[utf8]{inputenc}
\usepackage{graphicx}% Include figure files
\usepackage{dcolumn}% Align table columns on decimal point
\usepackage{amsmath}
\usepackage{bm}% bold math
\usepackage{ulem}
\usepackage[usenames,dvipsnames,svgnames,table]{xcolor}
\usepackage{braket}

\newcommand{\refeqn}[1]{(\ref{#1})} %Reference for equations
\newcommand{\us}{\,} %Defines the unit spacing
\newcommand{\Hamil}{\mathcal{H}} %Defines the typesetting for the Hamilton operator
\newcommand{\vect}[1]{\textbf #1}
\renewcommand{\exp}[1]{\text{e}^{#1}}

\begin{document}

%\preprint{APS/123-QED}

\title{Majorana bound state localization and energy oscillations for magnetic impurity chains on conventional superconductors}

\author{Andreas Theiler}
 \affiliation{
 Department of Physics and Astronomy, Uppsala University, Box 516, SE-751 20 Uppsala, Sweden
}
\author{Kristofer Bj\"ornson}
\affiliation{
 Department of Physics and Astronomy, Uppsala University, Box 516, SE-751 20 Uppsala, Sweden
}
\affiliation{
Niels Bohr Institute, University of Copenhagen, Juliane Maries Veg 30, DK-2100 Copenhagen, Denmark
}
\author{Annica M. Black-Schaffer}
\affiliation{
 Department of Physics and Astronomy, Uppsala University, Box 516, SE-751 20 Uppsala, Sweden
}%
\email{annica.black-schaffer@physics.uu.se}

\date{\today}% It is always \today, today,

\begin{abstract}
We study a chain of magnetic impurities on a conventional superconductor with spin-orbit coupling, treating the superconducting order fully self-consistently. 
We find and quantify strong hybridization between the topological edge Majorana bound states (MBSs) and in-gap Yu-Shiba-Rusinov (YSR) states, which causes increasing energy oscillations as a function of magnetic impurity strength, even when the direct MBS overlap is negligible.
By treating the MBS as a topological boundary state, dependent only on the effective mass gap, we arrive at a fully parameter-free functional form of the its localization which decreases with magnetic impurity strength, opposite to the behavior of the superconducting coherence length.
\end{abstract}

\maketitle

% --------------------------- %
% INTRODUCTION:
% --------------------------- %
\section{Introduction}
Single Majorana bound states (MBSs) at zero energy form at each end points of certain one-dimensional (1D) topological superconductors.\cite{kitaev2001unpaired,hasan2010colloquium,qi2011topological}
A MBS quasiparticle is its own antiparticle, and only by combining two MBSs one electronic degree of freedom is formed.\cite{wilczek2009majorana, Elliott&Franz2015} This inherent non-locality is exotic in itself and MBSs are also promising as building blocks for robust quantum computation.\cite{SternLindner2013,SarmaNPG2015}

Systems hosting MBSs typically combine superconductivity with spin-orbit coupling and magnetism. \cite{aguado2017majorana}
A much studied setup consist of a 1D chain of magnetic impurities on the surface of a conventional superconductor with an effective spin-orbit coupling\cite{Choy2011, nadj2013proposal, Brydon2015, Hui2015, vcadevz2016zero} or are closely related.\cite{sacramento2015magnetic, schecter2016self, christensen2016spiral, mashkoori2019majorana}
Experimental realizations, using e.g.~Fe impurities on a Pb surface, have measured zero-energy states very localized at the chain end points,\cite{nadj2014observation, pawlak2015probing,ruby2015end, feldman2016high} with the spin-polarization being consistent with MBSs and not other in-gap states,\cite{Jeoneaan3670} such as  Yu-Shiba-Rusinov (YSR) states, which are always present for magnetic impurities in superconductors.\cite{yu1965bound,shiba1968classical,rusinov1968superconcductivity}  

Intriguingly, the spatial extent of the measured zero-energy states is magnitudes smaller than the superconducting coherence length $\xi = \hbar v_F/\Delta$, where $v_F$ is the Fermi velocity and $\Delta$ the bulk superconducting order parameter, which usually sets the length scale in superconductors.
This discrepancy is also present in numerical studies.\cite{Klinovaja2012loc, nadj2014observation, Li2014, peng2015strong, PoyonenPRB2016, Mishmash2016}
One explanation put forward is that $\xi$ is renormalized on the chain to be much closer to experimental values,\cite{peng2015strong, Mishmash2016} with an additional suppressing power-law prefactor, due to the 2D environment.\cite{nadj2014observation,Li2014, Zyuzin2013,PoyonenPRB2016}
Strong localization, dependent on $\Delta$, has also been derived in the dilute impurity limit.\cite{pientka2013topological, Pientka2014Shibabands}

At the same time, magnetic impurities have for a long time been known to heavily suppress superconductivity at the impurity sites,\cite{balatsky2006impurity} even to the extent of producing a local $\pi$-shift in $\Delta$.\cite{Flatte97, Salkola97, Meng15, bjornson2017superconducting} Thus, properly allowing the superconducting order to respond to magnetic impurities always results in a nearly diverging $\xi$.
An alternative explanation to MBS localization, going beyond a (renormalized) $\xi$, is clearly needed in order to fully understand MBSs localization properties.

In this work we study a simple yet general model of a ferromagnetic impurity chain embedded in a 2D superconductor, capturing the qualitative behavior in both the dilute impurity and dense quantum wire limits.
We solve fully self-consistently for the superconducting order parameter, resulting in a strong suppression of $\Delta$ close to the chain.
Most importantly, we find that the lowest energy state in the topological phase is actually not just the topological edge state, the only state guaranteed at zero energy with Majorana non-Abelian statistics and thus the only MBS, but it also inherits significant character from YSR states. 
The strong hybridization between the zero-energy MBS and YSR states directly explains the large energy oscillations in the lowest energy state, growing with increasing magnetic impurity strength. Moreover, by treating the MBS as a topological boundary mode with its wave function determined by the effective mass gap, we arrive at a parameter-free simple functional fit for the MBS localization length showing good agreement with our numerical results. 
Notably, we find that the MBS localization length decreases with increasing magnetic impurity strength.
Taken together, these results provide a unifying picture of MBS interactions, localization, and energy oscillations.

% --------------------------- %
% SETUP:
% --------------------------- %
\section{Model and method}
We study a chain of ferromagnetically aligned impurities on a conventional $s$-wave superconductor surface with Rashba spin-orbit coupling.
The simplest Hamiltonian to fully describe this system is $\Hamil = \Hamil_{0} + \Hamil_{\textrm{sc}}  + \Hamil_{\textrm{im}}$,\cite{sato2009non, sato2010non, Sau2010, black2011majorana, bjornson2013vortex,nadj2013proposal,pershoguba2015currents, bjornson2015probing, bjoernson2015spin, bjornson2016majorana} where
\begin{align}
\label{eqn:Model_Hamiltonian}
 &\Hamil_{0} = \sum_{\vect{i},\vect{j},\sigma} t_{\vect{i},\vect{j}} c^\dagger_{\vect{i}\sigma} c_{\vect{j}\sigma}  + 
\alpha \sum_{\vect{i},\vect{b}} \exp{i \theta_\vect{b}} c^\dagger_{\vect{i}+\vect{b} \downarrow} c_{\vect{i} \uparrow} + \textrm{H.c.}, \nonumber \\
&\Hamil_{\textrm{sc}} = \sum_{\vect{i}} \Delta_\vect{i} c^\dagger_{\vect{i}\uparrow} c^\dagger_{\vect{i}\downarrow} + \textrm{H.c.}, \,
\Hamil_{\textrm{im}} = \! \sum_{\vect{a}, \sigma, \sigma'} \!\! \!-V_z\left(\vect{a}\right) c^\dagger_{\vect{a}\sigma} \sigma^{z}_{\sigma \sigma'} c_{\vect{a}\sigma'} .
\end{align}
\begin{figure}[thb]
    \includegraphics[width=0.4\textwidth]{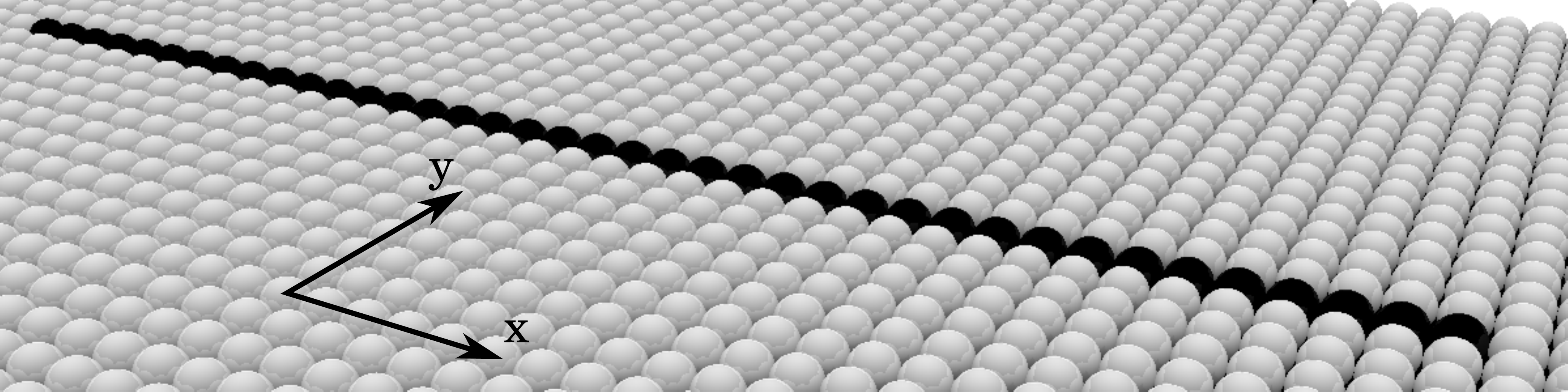}
	\caption{
	Sketch of the 2D setup. White lattice sites represent a $s$-wave superconductor with Rashba-like spin-orbit coupling. Black impurity sites are magnetic impurity sites which include an additional Zeeman interaction term pointing out of plane.}
	\label{fig:setup_sketch}
\end{figure}
Here $\Hamil_{0} + \Hamil_{\textrm{sc}}$ describes a 2D square lattice superconductor, with nearest neighbor hopping $t_{\vect{i},\vect{i}+\vect{b}}=-t$, chemical potential $t_{\vect{i},\vect{i}}=\mu$, and spin-orbit coupling $\alpha$, with the polar angle $\theta_\vect{b}$ formed by the nearest neighbor bond $\vect{b}$. Superconductivity is given by the site-dependent $s$-wave order parameter $\Delta_{\vect{i}}$. The magnetic impurities are captured by $\Hamil_{\textrm{im}}$ as classical spins aligned out-of-plane with strength $V_z$ and only present on sites $\vect{a}$, forming a 1D chain, well surrounded by superconducting sites, see Fig.~\ref{fig:setup_sketch}.  

Each magnetic impurity produces two YSR states in the superconductor, which for a chain merge into 1D so-called Shiba bands.
The Hamiltonian Eq.~\eqref{eqn:Model_Hamiltonian} thus describes an effective 1D Shiba band model.\cite{pientka2013topological,Pientka2014Shibabands,Weststrom2015}
Note that the model lattice constant $a$ does not have to be the physical lattice constant, but can be thought of as a larger course-graining distance, such that the magnetic impurities are in a more dilute limit. 
Alternatively, Eq.~\eqref{eqn:Model_Hamiltonian} can equally well be viewed as a partly polarized (unless $V_z$ is large) ferromagnetic quantum wire with spin-orbit coupling that is embedded in a void and tunnel-coupled to the surrounding superconductor. This model represents magnetic impurities in the quantum wire or hybridizing dense limit.\cite{klinovaja2013topological, braunecker2013interplay,vazifeh2013self} \footnote{Since self-consistency gives $\Delta \approx 0$ on all impurity sites, a finite superconducting pairing potential $U$ on these sites does not change the results.}
By using the same kinetic energy parameter $t$ for both the superconductor and chain sites, as well as for the coupling between superconductor and chain, we thus capture within a single simple model qualitatively both the Shiba band and the quantum wire scenarios, which are the two limits frequently discussed for topological magnetic impurity chains.\cite{Hoffman2016, Andolina2017}
While adding further parameters can make the model more tuned to a specific experimental realization, the major benefit of Eq.~\eqref{eqn:Model_Hamiltonian} is its simplicity and generality.
Moreover, adding an explicit $p$-wave order parameter due to intrinsic spin-orbit coupling has been shown to not change the results.\cite{Bjornson2017island}

We solve Eq.~\eqref{eqn:Model_Hamiltonian} using the Bogoliubov-de Gennes (BdG) formalism.\cite{gennes1999}, treating the superconducting order parameter fully self-consistently.\cite{BlackSchaffer2008, bjornson2013vortex,Reis2014,bjornson2015probing, Awoga2017}
We thus only assume that the superconductor provides a propensity for electron pairing, which we model with a constant on-site attraction $U$ to emulate conventional spin-singlet $s$-wave pairing.
The order parameter  $\Delta_{\bf i}$ is then calculated on each site of the lattice through the self-consistent condition
\begin{equation}
    \label{eqn:delta_definition}
            \Delta_{\vect{i}} = -U\left\langle c_{\vect{i}\downarrow} c_{\vect{i}\uparrow} \right\rangle \\
            = - U \! \sum_{\nu, E_{\nu} < 0} u_{\nu \vect{i} \uparrow} v_{\nu \vect{i} \downarrow}^*,
\end{equation}
where $ u_{\nu \vect{i} \uparrow}$ and $v_{\nu \vect{i} \downarrow} $ are the electron and hole components of the eigenstate of the Hamiltonian Eq.~\eqref{eqn:Model_Hamiltonian}, with up and down spin at site $\vect{i}$, respectively.
The sum in Eq.~\eqref{eqn:delta_definition} is over all states below the Fermi energy.
Starting with an initial guess for $\Delta_\vect{i}$, we iteratively calculate new $\Delta_{\vect{i}}$'s until the maximal local change relative to $\Delta_{bulk}$ of two subsequent iterations becomes negligible small ($\leq 10^{-4}$).

We study system sizes up to $80 \times 41$  sites  and here report results for $L = 40$, $\alpha = 0.3\us t$, $\mu = -4 \us t$, and $U$ such that $\Delta = 0.3 \us t$ in the bulk.
These choices make the superconductor metallic in the normal state and allow access the topological phase at smallest $V_z$ since in the bulk the topological phase transition (TPT) is at $V_c = \pm \sqrt{\Delta^2 + (4t-\mu)^2}$.\cite{sato2010non}
We have verified that the results are not sensitive to these parameter choices, as long as we stay within the same topological phase.
To calculate the self-consistent $\Delta_{\vect{i}}$ profile for this system we use a Chebyshev polynomial method\cite{weisse2006kernel, covaci2010efficient} to expand the Green’s functions non-principal part, using up to 10000 Chebyshev coefficients.
We access wave functions and their energies using Arnoldi iteration with the self-consistent $\Delta$ solution.
The calculations are implemented using the TBTK software development kit.\cite{TBTK, kristofer_bjornson_2017_556398}

% --------------------------- %
% SELF-CONSISTENCY
% --------------------------- %
\section{Effects of Self-consistency}
\begin{figure}[thb]
%	\subfigimg[width=0.47\textwidth]{\vspace*{0.3cm} \hspace*{-0.8cm}(a)}{pictures/delta_profile.pdf}
%    \subfigimg[width=0.47\textwidth]{\vspace*{0.1cm} \hspace*{-0.8cm}(b)}{pictures/energy_spectrum_only2}
    \includegraphics[width=0.47\textwidth]{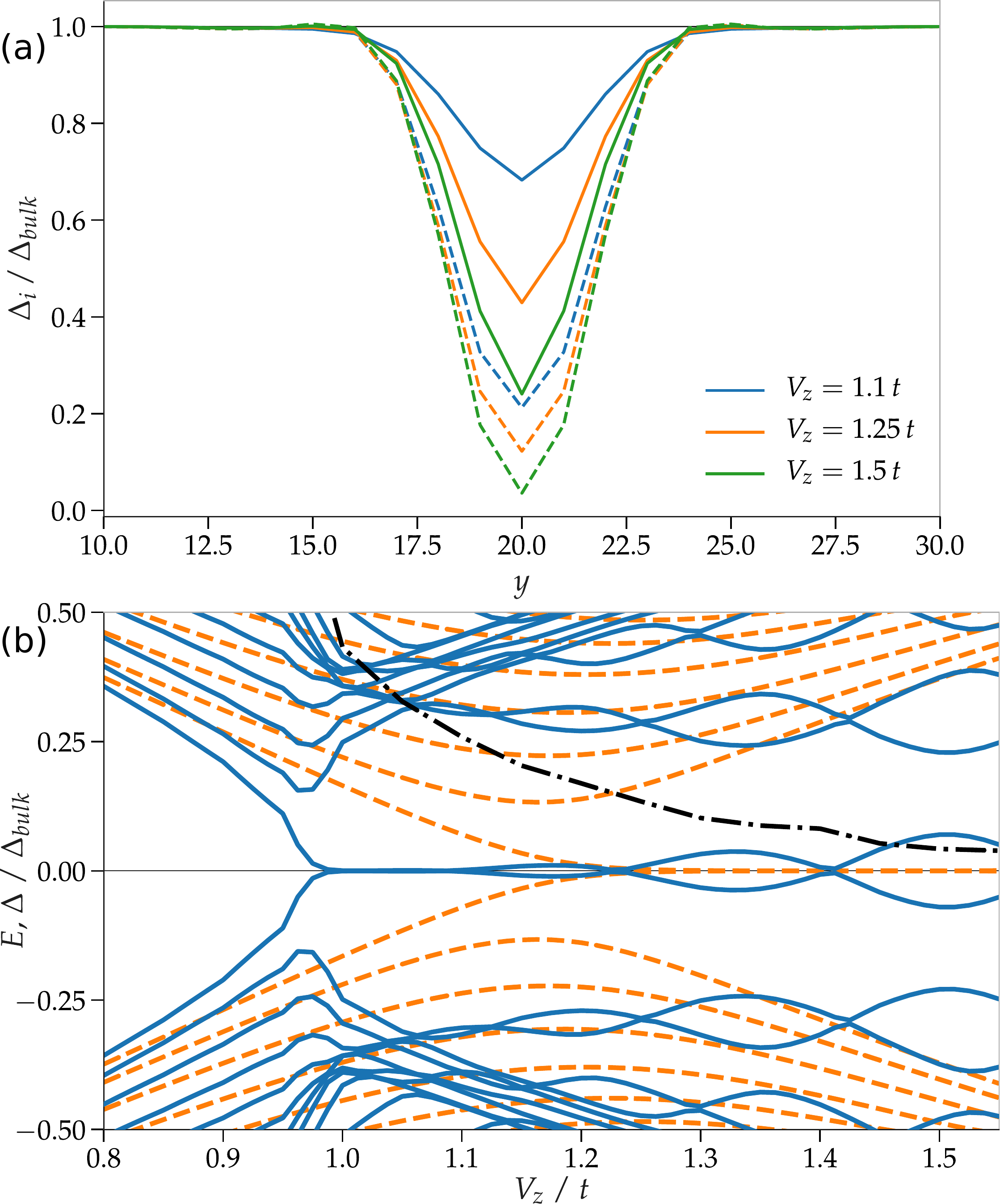}
	\caption{
	(a) Self-consistent order parameter profile $\Delta_{i}$ extracted along a direction $y$ perpendicular to the chain cutting through the middle (dashed) and at the chain end points (solid) for several values of $V_z$. (b) Low-energy spectrum as a function of $V_z$ for self-consistent (solid blue) and non-self-consistent (dashed orange) solutions and average $\Delta$ along chain (dash-dotted black).}
	\label{fig:selfcons}
\end{figure}

We first establish the importance of a self-consistent solution for the superconducting order.
The magnetic impurities dramatically suppress $\Delta$ locally around the chain because of the local pair breaking effect, with the length scale of this phenomena being typically set by the Fermi wave vector.\cite{Flatte97}
This is visualized in Fig.~\ref{fig:selfcons}(a), where we plot $\Delta_{\vect{i}}$ across the chain both in the middle and at the end of the chain, as well as in Fig.~\ref{fig:selfcons}(b) where we plot the average over all chain sites as a function of $V_z$ (b, black line).
As seen, the suppression is site-dependent along the chain, showing somewhat larger suppression in the middle as compared to the end points.
We here note that the highly localized suppression of the order parameter makes our choice of a 2D superconductor sufficient. Including also the third dimension for a bulk superconductor would only result in slightly modified parameters. Also, since the magnetic impurities are located on neighboring sites, only their short-range interactions are important. Thus, any the dimensional differences in the long-range decay of YSR states are here not important \cite{rusinov1968superconcductivity}. 
%
% FIGURE 1:

The dramatic $\Delta$ suppression on the chain sites has large consequences for both the energy spectrum and the TPT.
In Fig.\ref{fig:selfcons}(b) we plot the full Hamiltonian Eq.~\eqref{eqn:Model_Hamiltonian} low-energy spectrum for the self-consistent solution (blue) and without self-consistency (orange).
In the latter case $\Delta_{\vect{i}} =0.3t$, equaling the bulk value.
Most notably, the TPT, where the MBS appear, takes place at significantly lower $V_z$ in the self-consistent solution.
Since the TPT in a homogeneous system occurs at $V_z > \Delta$ (for $\mu = -4t$), this can be understood as a direct consequence of $\Delta$ being heavily suppressed on the chain sites.
Thus, the chain enters an effective 1D topological phase at a lower $V_z$, despite being fully embedded in a large superconductor. The TPT is also more distinct in the self-consistent case, with the energy levels dropping more sharply to zero. The self-consistent TPT for a short chain thus resembles the situation in the infinite bulk much more than has previously been suggested.\cite{Mishmash2016} This should make for experimentally easier detection, even for short chains.

% --------------------------- %
% MBS INTERACTIONS
% --------------------------- %
\section{MBS interactions}
Next we turn our attention to the oscillations in the lowest energy state, henceforth $\psi_0$, especially prominent with increasing $V_z$ as seen in Fig.~\ref{fig:selfcons}(b). We note that these oscillations are also present in the non-self-consistent solution, but only at larger $V_z$ as the TPT appears at larger $V_z$.
According to conventional wisdom these oscillations arise due to spatial overlap of the two end-point MBSs.\cite{DasSarma2012,Prada2012, nadj2014observation, pientka2013topological, pawlak2015probing, Cayao2017osccurr, Andolina2017}
However, the state guaranteed at zero energy, and hence a MBS, is the topological boundary state, which can be expressed as the generic edge state solution:\cite{hasan2010colloquium,qi2011topological} 
\begin{equation}
\label{eqn:generic_edge_state}
    \varphi_{M}(x) = C \exp{\frac{1}{\alpha} \int^x_0 M\left(x'\right) dx'},
\end{equation}
where $M(x')$ is the system's mass gap.
In an ideal 1D model, i.e.~keeping only the magnetic chain, the mass gap is simply given by $M = |\Delta| - |V_z|$ (for $\mu =-4t$).\cite{sato2010non, bjornson2015probing}
The mass gap thus changes sign at the TPT and then increases to larger negative values as we further increase $V_z$. Thus, the topological origin of the MBS guarantees stronger localization with increasing $V_z$, which should give diminishing energy oscillations; exactly contrary to the behavior of $\psi_0$. 
To solve this conundrum, we are forced to interpret the lowest energy state $\psi_0$ in the topological phase as not just the zero-energy topological edge state, i.e.~the MBS, but also containing significant contributions from other states.
Looking critically at the energy spectrum in Fig.~\ref{fig:selfcons}(b) this is actually not surprising as it can be viewed as multiple avoided crossings between different states.
Consequently, the energy oscillations in $\psi_0$ are not due to direct MBS-MBS interaction, but are the result of interactions between the MBS and YSR subgap states.

\begin{figure*}[thb]
	\includegraphics[width=\textwidth]{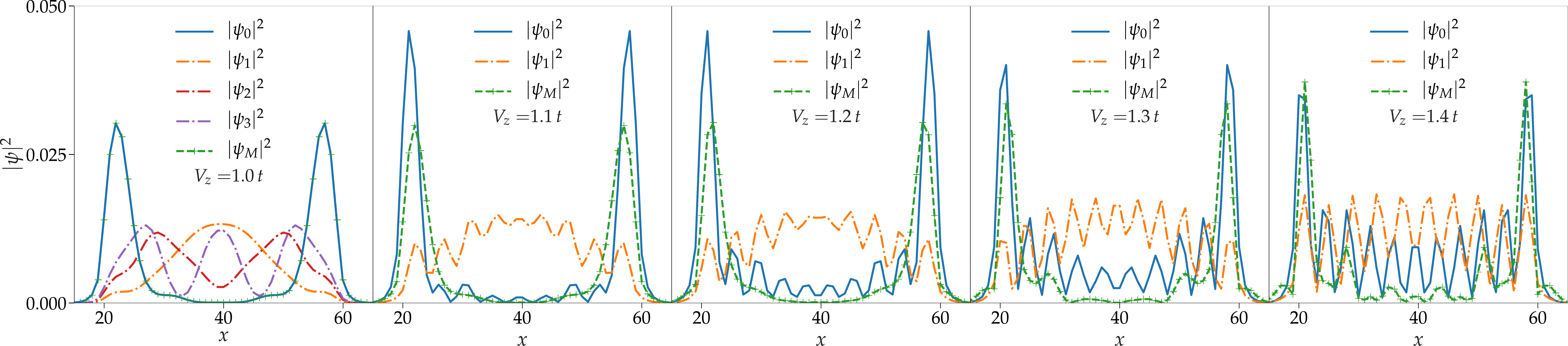}
	\caption{Densities of lowest energy state $\psi_{0}$ (solid), higher energy states $\psi_{1,2,3}$ (dash-dotted), and projected MBS $\psi_M$ from Eq.~\eqref{eqn:corrected_mbs} (dotted) along the chain for various values of $V_z$.
    }
	\label{fig:wave_functions}
\end{figure*}

The above interpretation is clearly illustrated in Fig.~\ref{fig:wave_functions}, where we show the densities of the the wave functions for the lowest lying energy states $\psi_i$, at $V_z$ values ranging from $V_z =1$ to  $1.4 \us t$.
At $V_z = 1 \us t$, i.e.~close to the TPT, the lowest energy states $\psi_0$ is localized at the chain ends, thus having mainly topological MBS character. 
Moreover, the higher energy states, $\psi_{1,2,3}$, are clearly forming standing waves along the chain, with wave numbers rising with energy, such that $\psi_0$ and these lowest lying YSR band states do not have much spatial overlap. 
As $V_z$ is increased these YSR states move subsequently down in energy towards the Fermi level.
If it was not for the finite Rashba spin-orbit interaction, the states would even cross the Fermi level.\cite{bjornson2017superconducting}
Consequently, the number of nodes in the lowest energy YSR state increases with $V_z$, as also clearly illustrated in Fig.~\ref{fig:wave_functions}.
Since YSR states with more nodes have larger weights at the wire end points, this leads to a stronger hybridization with the MBS at larger $V_z$ and then automatically larger energy oscillations. 
The increasing hybridization between MBS and YSR states is also clearly visible in that the overall eigenstate shapes  become more similar.
For example, in Fig.~\ref{fig:wave_functions}, the overall shape of the lowest and first excited states show very clear similarities at $V_z = 1.1$ and $V_z = 1.2 \us t$.

\begin{figure}[htb]
        \includegraphics[width=0.47\textwidth]{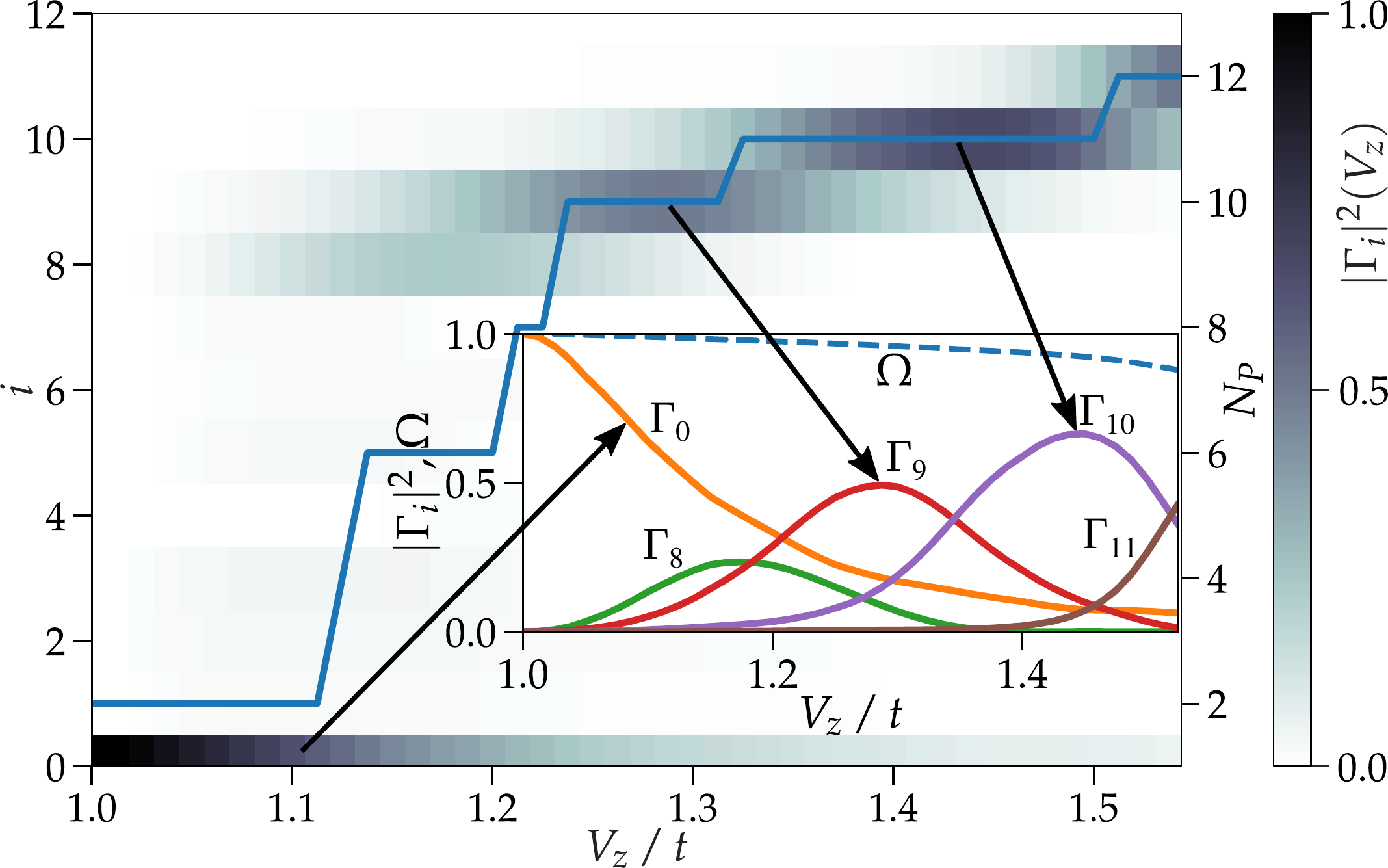}
	\caption{Overlap densities $|\Gamma_i|^2$ (greyscale) as a function of index $i$ (left $y$-axis) and $V_z$. Peak number $N_P$ in $\psi_0$ along the $x$-axis (solid line, right $y$-axis).
    The two $y$-axes are aligned such that the number of peaks in $\tilde{\psi}_{i}$ align with the correct $i$ index for $i = 8$ to 11.\cite{peakFootnote}
    Inset shows complementary picture for the largest $\Gamma_i$, while dashed line denotes the completeness $\Omega$ of the basis states.
    }
	\label{fig:mediating_states}
\end{figure}

Having illustrated the strong interaction between MBS and YSR states, we now quantify these arguments. Specifically, we need to project the lowest energy state $\psi_0$ onto the true topological boundary state, the MBS, and the clean, i.e.~non-hybridized, YSR states.
However, we do not have easy access to neither the exact MBS nor the clean YSR states, as we only numerically have the orthogonal eigenstates which includes all hybridizations.
Thankfully, we find in Figs.~\ref{fig:selfcons}(b) and \ref{fig:wave_functions} (for $V_z = 1.0 \us t$), that at the TPT $\psi_0$ is essentially the clean topological boundary state and the higher energy states are the YSR states with only negligible hybridization.
Thus we can quantify the amount of MBS and YSR character in the lowest energy state $\psi_0$ by projecting it on a basis spanned by the subgap states (energies $|E_i| < \Delta$, $\sim 0.2\%$ of the available Hilbert space) just past the TPT:
\begin{equation}
\label{eqn:mbs_decomposition}
 \ket{\psi_{0}(V_z)} \approx \sum_{i} \Gamma_i(V_z) \ket{\tilde{\psi}_i},
\end{equation}
%
%TODO check this statement (which delta).
where $\tilde{\psi}_i = \psi_i(V_z \approx V_c)$ and $i$ indexes the states by increasing energy.
The overlap coefficients $\Gamma_i(V_z) = \braket{\psi_0(V_z)|\tilde{\psi}_i}$ 
measure the MBS ($\tilde{\psi}_0$) and YSR ($\tilde{\psi}_{i\neq 0}$) components of $\psi(V_z)$ \footnote{The numerics give a symmetrical and anti-symmetrical solution when combining the $\psi_0$ state above zero energy with its partner below zero energy. Thus, the overlaps $\Gamma_i$ with even- or odd-parity YSR states are interchangeably zero.
To avoid this behavior, Fig.~\ref{fig:mediating_states} shows the solution for both $\psi_0$ states simultaneously.}.
Hence, a large $\Gamma_0(V_z)$ indicates that the $\psi_0$ state is essentially the true topological boundary state, i.e.~the MBS, while if $\Gamma_{i \neq 0}$ increases it shows that $\psi_0$ contains contributions from the $i^{\text{th}}$ YSR state. 

Fig.~\ref{fig:mediating_states} shows $|\Gamma_i|^2$ in greyscale for increasing $V_z$, with the inset offering a complementary line plot for the most significant $\Gamma_i$. As seen, $\Gamma_0$ is large close to the TPT but then significantly drops, where instead $\Gamma_{i}$ for $i \geq 8$ becomes large, which is the overlap with the YSR state at lowest energy.
Moreover, there exists a direct correspondence between the largest $\Gamma_i$ component and the number of peaks $N_P$ in $\psi_0$, as illustrated by the blue line.
Here a peak is numerically defined as maxima with a difference of $ >10 \us \%$ in value to the surrounding minima.
$N_P$ tracks closely the dominant $\Gamma_i$ coefficient, providing further evidence of hybridization between the lowest energy YSR state and the MBS.
Taken together, this proves that the lowest-energy state, $\psi_0$, is not just the MBS but also host significant character from the lowest-energy YSR state. Notably, YSR states oscillate proportional to $\text{cos} (k_F x)$,\cite{yu1965bound, shiba1968classical, rusinov1968superconcductivity} with $k_F$ the Fermi momentum, and gives the resulting state it's oscillating tail while also mediating the interaction between the two edge states to split their energy.

To validate our results, we quantify the completeness of the basis in Eq.~\eqref{eqn:mbs_decomposition} by calculating
\begin{equation}
    \label{eqn:basis_completness}
    \Omega(V_z) = \sum_i |\braket{\psi_0(V_z)|\tilde{\psi}_i}|^2,
\end{equation}
which is plotted in the inset in Fig.~\ref{fig:mediating_states}.
As seen, $\Omega$ only drops to  about $0.9$ at high $V_z$ values, and thus the chosen basis captures the physics of $\psi_0$ over a wide $V_z$ range.
The drop at higher $V_z$ can be explained by the intrinsic change of $\varphi_M (x)$ due to the increased mass gap in Eq.~\refeqn{eqn:generic_edge_state}.

We also rearrange Eqn.~\eqref{eqn:mbs_decomposition} to arrive at a numerical approximation for the topological MBS by subtracting all contributions from the YSR interactions:
\begin{equation}
\label{eqn:corrected_mbs}
 \ket{\psi_M(V_{z})} \approx \ket{\psi_{0}(V_z)} - \sum_{i \neq 0} \Gamma_i(V_z)  \ket{\tilde{\psi}_i},
\end{equation}
resulting in a smooth wave function without any additional peaks as seen in Fig.~\ref{fig:wave_functions}.
This true topological MBS provides a good agreement with the first $\psi_0$ peak, especially for values of $V_z$ up to $1.25 \us t$.
In this range our way of extracting the true topological MBS works very well, with almost no signatures of interactions with YSR left, as evidenced in the lack of oscillations in $\psi_M$.
At higher values some weaker ringing is still visible in $\psi_M$ in the interior of the chain. This can be attributed to the loss of completeness in the basis states $\Omega$.

% --------------------------- %
% MBS LOCALIZATION
% --------------------------- %
\section{MBS localization}
\begin{figure}[thb]
\begin{center}

	\includegraphics[width=220pt]{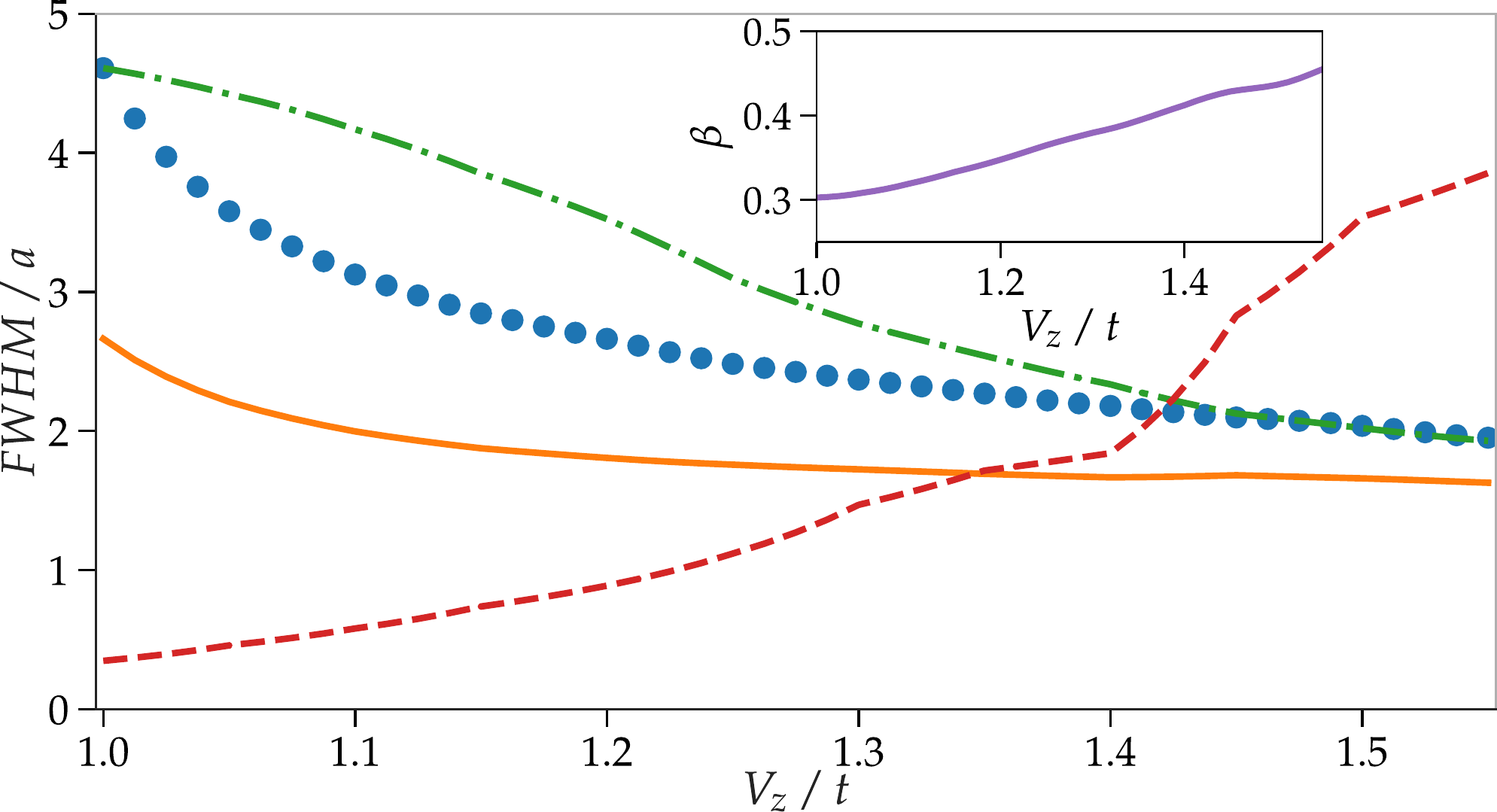} %previous width = 230pt
	\caption{
	 MBS FWHM at chain end point as a function of $V_z$ extracted from data $\psi_0$ (dots), numerical approximation $\psi_M$ (dash dotted), and fit using $\varphi_{M}$ (solid). Compared to FWHM based on $\xi$ (dashed). Inset: $\beta$ calculated as the ratio of $|\psi_{0}|^2$ located on the chain.
}
	\label{fig:local}
\end{center}
\end{figure}

Having found a fully consistent interpretation of both wave function and energy spectrum behavior, we turn to the MBS localization length.
Numerically, we find this as the full width at half maximum (FWHM) of $\psi_M$ in Eq.~\eqref{eqn:corrected_mbs} or, as we shown above, equivalently, of the $\psi_0$ edge peak.
Alternatively, we can access the functional form of the localization from the ideal edge state properties of $\varphi_M (x)$ through Eq.~\refeqn{eqn:generic_edge_state}.
In a 1D model the mass gap $M = |\Delta| - |V_z|$ is a constant.  
However, for a 1D chain embedded in a superconductor this is no longer true.
First, $\Delta$ depends on the position $x$ along the chain and is heavily suppressed compared to the bulk $\Delta$.
Second, $V_z$ is only finite on the chain sites, but zero elsewhere.
The effectiveness of $V_z$ thus becomes diluted, since both the MBSs and YSR states are spread out over a small but finite region orthogonal to the chain.
We take both of these effects into account by setting the mass gap to $M(x) = |\Delta(x)| - \beta|V_z(x)|$, where $\Delta(x)$ is determined self-consistently at each $x$ and the dilution effect of $V_z$, i.e.~$\beta$, is the fraction of the MBS state that is localized on the chain sites. 
The inset of Fig.~\ref{fig:local} shows how $\beta$ increases somewhat with $V_z$, indicating an increasing localization on the chain.
We also use constant $\beta$'s and find the same qualitative trend for the MBS localization length for all $\beta \in [0.1, 0.5]$.
With these modifications, we only have to fix the overall constant $C$ in $\varphi_M$ to the total height of the $\psi_0$ boundary peak, to achieve a {\it fully parameter-free} functional form of the MBS wave function.

In Fig.~\ref{fig:local} we compare the numerically extracted FWHM of the self-consistent $\psi_0$ MBS peak (dots) and $\psi_M$ (dash-dotted), with the prediction $\varphi_M$ FWHM (solid).
Not only do we produce the same order of magnitude for the FWHM in all cases, but also accurately capture how the localization length clearly decreases with increasing field $V_z$, since the mass gap increases approximately linearly with $V_z$. The clear agreement between the curves shows that $\psi_M$ is the true topological edge state.
%
% FIGURE 3:
To contrast our purely topology-based approach to the MBS localization, we also estimate the localization based on the superconducting coherence length $\xi$.
To account for self-consistency we use $\xi = C'/\Delta$, where $\Delta$ is the average order parameter on the the chain and $C'$ an overall constant.
We determine $C'$ such that the FWHM fits the numerical results.
However, no matter the exact details, the MBS localization length based on fits involving $\xi$ always increases with $V_z$ (dashed line).
Even if the self-consistent suppression of $\Delta$ is ignored, the MBS localization length would only be a constant function of $V_z$.
We therefore conclude that the MBS localization is not related to the (renormalized) $\xi$, though renormalization can still be important.

% --------------------------- %
% Conclusions:
% --------------------------- %
\section{Concluding remarks}
Using a self-consistent treatment of superconductivity we show how the lowest energy state in the topological phase does not only contain the topological boundary state, i.e.~the MBS, but also incorporates significant character from YSR states. This results in the lowest energy state having both a strongly modified wave function away from the ideal MBS shape and large energy oscillations. By numerically extracting the YSR state components we can however isolate the MBS and we show that it forms a single well-localized and non-oscillating peak at the chain end point. We find that the MBS localization length  decreases with $V_z$ and is only a function of the effective mass gap, and thus not governed by the superconducting coherence length.

Beyond providing a unifying framework for MBS interactions, localization, and energy oscillations, these results importantly give valuable insight in how to engineer systems with cleaner features, where the lowest energy state also has more MBS character and less YSR contributions.
For example, longer impurity chains result in more extended YSR states, directly leading to less interactions with the MBSs and therefore more MBS contributions in the lowest energy state. Also disorder can decrease the MBS interaction by localizing the YSR states, which is consistent with a remarkable MBS disorder robustness.\cite{Awoga2017}
Additionally, we speculate that Coulomb repulsion may push the YSR states to higher energies, offering yet another way to decrease MBS interactions.

\acknowledgments
We thank J.~Klinovaja, D.~Loss, and F.~von Oppen for initial discussions related to this project. This work was supported by the Swedish Research Council (Vetenskapsr\aa det) Grant No.~621-2014-3721, the Swedish Foundation for Strategic Research (SSF), and the Wallenberg Academy Fellows program through the Knut and Alice Wallenberg Foundation.


\begin{thebibliography}{68}
\bibitem [{pea()}]{peakFootnote}%
  \BibitemOpen
  \href@noop {} {}\bibinfo {note} {For non-interacting YSR states, the index
  $i$ coincides directly with the number of peaks in $|\psi_i|^2$ along the
  wire. However, at the TPT, some of the YSR states have already hybridized and
  this rule is no longer valid. Through inspection we instead find that the
  states with indices $i = 8$ to 11 have $i + 1$ peaks.}\BibitemShut {Stop}%
\end{thebibliography}
\end{document}